\documentclass[sigconf, authorversion, nonacm]{acmart}

\AtBeginDocument{%
  \providecommand\BibTeX{{%
    \normalfont B\kern-0.5em{\scshape i\kern-0.25em b}\kern-0.8em\TeX}}}

\setcopyright{none}

\begin{document}

\title{Is it Fun?: Understanding Enjoyment in Non-Game HCI Research}


\author{Michinari Kono}
\affiliation{%
  \institution{Bandai Namco Research Inc.}
  \streetaddress{Koto-ku}
  \city{Tokyo}
  \country{Japan}}
\email{m2-kono@bandainamco-mirai.com}

\author{Koichi Araake}
\affiliation{%
  \institution{Bandai Namco Research Inc.}
  \streetaddress{Koto-ku}
  \city{Tokyo}
  \country{Japan}}
\email{k-araake@bandainamco-mirai.com}









\begin{abstract}
An experience of {\it fun} can be an important factor for validating the value of games. In several studies, research on non-game human–computer interaction (HCI) has been attempted to measure the enjoyment of work. However, a majority of the studies do not discuss the importance and value of the result. It is not clear as to how the term {\it fun} is understood in a non-game context nowadays. To analyze this shortcoming, we reviewed extant studies, particularly studies on non-game HCI, and explored as to how researchers determine if the value of an activity is fun. Consequently, we discussed and categorized the usage of the terms and analyzed the methodologies that are used in extant studies that evaluate the effects of fun and related terms. To gain a better understanding of fun in HCI, we provided several directions that can be discussed for strengthening enjoyable HCI research beyond applications involving games.
\end{abstract}




\maketitle

\section{Introduction}
{\it Fun} or {\it Enjoyment} can be considered as key factors in developing successful games. In the field of games and play, researchers of human-computer interaction (HCI) work on computer games at conferences such as Conference on Human Factors in Computing Systems (CHI) and Symposium on Computer-Human Interaction in Play (CHI Play). Researchers and game designers have been working on untangling the role and paradigms of games. Specifically, ``enjoyment'' is often stated as a core value and objective for people playing game~\cite{10.5555/772072.772128, mellecker2013disentangling}. 
However, researchers are measuring, reporting, and arguing about indicators of fun in studies related to non-game and non-play HCI. Although several researchers and designers have discussed the value of fun in recent games and play research, it is unclear as to how an activity being {\it fun} can be effective in other cases in the recent CHI field. 

Back in 1988, Caroll and Thomas~\cite{fun1988} discussed the importance of fun along with the confusion of fun and usability. The importance of fun was introduced by referring to literature related to game mechanics, such as reward and adventure games. In their article, they questioned, ``{\it Would a computer scientist who tries to build a
professional career studying fun be taken seriously (pp. 23)?}'' and at the very end, they even said ``{\it We realize that many people will read this article as a sort of joke (pp. 23).}'' At that time, more than 30 years ago, it was quite difficult for us to imagine discussing non-game HCI work through the lens of {\it fun}. However, beyond their argument, HCI started to think about enjoyment more seriously in the 2000s~\cite{blythe2004funology, funsoftware, pleasureusability, uxagenda}. Although researchers have used fun or enjoyment as a metric for empirical studies in non-game research in HCI nowadays, there is a paucity of evidence on how it can be valuable in this context. Therefore, we address the following questions: How well do we understand the role of fun in HCI research that is not discussed in the scope of games and play? How are they validated? How are they discussed, and how can we improve them? The aim of this study is to discuss these questions and provide directions for addressing fun in HCI research. 

We hypothesize that many studies on non-game-related HCI under-evaluate the importance of an activity being fun. Even though a number of studies that do not focus on game topics attempt to measure the enjoyment of non-game-related HCI. However, a majority of the studies do not measure enjoyment and fail to discuss the importance and value of the result. Conversely, for game research, many thorough discussions have been conducted. Enjoyment is considered as a fundamental value and is a successful area that already understands the role of fun~\cite{enjoymentgames}. Our aim is to aid future non-game HCI researchers in understanding the values of enjoyment and to encourage them to report and discuss potential ideas that may arise due to enjoyment.

To answer the aforementioned questions and goals, we performed a systematic survey of recent CHI and CHI play proceedings in which terms related to fun were used in a non-game context. We then categorized these studies into several paradigms to show the tendencies of the cases. Given that there are several approaches to measuring fun in these studies, we performed an analysis of extant studies wherein fun is evaluated or used as a metric. Finally, we discussed the differences in perspectives with respect to fun in game research according to Mekler et al.~\cite{enjoymentgames}, and suggested potential areas where HCI researchers can find opportunities for discussing and using fun as a metric or evidence. 

It should be noted that fun and enjoyment will be used interchangeably hereafter (other related terms are introduced later in the paper). Although fun and enjoyment are used interchangeably, they are chosen based on extant studies or adjusted subjectively for readability. As noted in the searching procedure section, we started by searching in dictionaries to omit possible biases due to games and play knowledge (we are aware that they can be used in a slightly different context~\cite{semanticsoffun}). Without such knowledge, the term ``fun'' was described with the other introduced terms (e.g., pleasure). This implies that the term can be used without distinguishment, if without any biases, and can be understood as an objective fact. Therefore, before the literature review, our definition of fun will rely on the definitions introduced in dictionaries as a ``term.''

\section{Understanding Fun}
Fun can be intrinsic and extrinsic motivations for users (or players). These motivations have been discussed from various perspectives and applied to several cases. In this section, we summarize related concepts of fun and enjoyment to better understand their current roles.

\subsection{Related Concepts}
Fun or enjoyment is considered as an important element in games and play. In 2014, Carter et al. studied papers from CHI and argued that game research in HCI can be categorized into four distinct paradigms~\cite{gamesinhci}: operative, epistemological, ontological, and practice. In {\it Man, Play, and Games}~\cite{manplaygames}, six definitions, including freedom and uncertainty, were discussed. In {\it Rules of Play}~\cite{rulesofplay}, the correlation between games and play was explained wherein games were a subset of play and play was a component of games. Juul further redefined games in the book {\it Half-Real}~\cite{halfreal}, and noted that a game can be enjoyable from different perspectives based on individuals, and it is difficult to explain this idea in a single statement. Vorderer et al.~\cite{enjoymententertainment} presented a conceptual model that considers enjoyment as a core component of entertainment. The motivation for entertainment can be described from three perspectives: escapism, mood management, and the wish to be challenged. Bernhaupt~\cite{evaluationentertainment} noted that HCI research and game research share methods to evaluate user experiences in games and entertainment. Fun is a concept that is evaluated in these types of studies. Related to the term fun, Tiger discussed the variations of pleasure and described it as an evolutionary entitlement~\cite{tiger1992pursuit}. Other concepts such as immersion were considered as important factors that should be measured in games~\cite{immersiongame}. 

The scope of behavioral change is an interesting concept with respect to play or engagement. Niedderer et al.~\cite{sustainabledfbc, dfbc} discussed the design of several behavioral change approaches. O'Brien and Toms~\cite{engageframe} presented a framework for the process of engagement. The process begins with engagement from factors such as aesthetics or interest, and leads to disengagement owing to factors such as usability and challenge. Furthermore, these factors share attributes with flow and play. Reguera et al.~\cite{quantifyengagement} presented a quantitative analysis of engagement dynamics into playful activities based on a popular video game. In these approaches, frameworks such as nudge~\cite{nudge} and the flow theory~\cite{flowtheory} are often referred to and discussed.

\subsection{Fun in Games}
Enjoyment is often discussed with factors such as player experience and usability~\cite{fourfun, enjoymentgames}. Mekler et al.~\cite{enjoymentgames} argued that enjoyment is central to player experience, and they conducted a systematic review based on game research based on the scope of quantitative studies on enjoyment. They explained that measurements with respect to flow were often noticed and in some studies, flow and enjoyment were used interchangeably~\cite{doi:10.1080/08838151.2011.546248}. Tyack and Mekler~\cite{selfdethcigame} further reviewed how the self-determination theory~\cite{selfdetermination} (a psychological theory of human motivation) was used in current HCI game research and found that the concepts of need satisfaction and intrinsic motivation are being applied widely for player experience, but other prominent core concepts are rarely considered. Lazzaro~\cite{fourfun} argued that usability is not the only component that can improve the quality of games and presented the four fun keys for games. Experience in games was described using four perspectives: hard fun, easy fun, serious fun, and people fun. They were determined as motivations for playing games. In summary, we play games for an opportunity to compete, explore a new experience, feel specific emotions, and spend time with friends. Furthermore, user experience or usability can be an important factor for research and design in a non-game context. Therefore, several concepts can be adopted from extant studies focusing on games for investigating the role of enjoyment.

{\it Gamification} is another highly related concept for discussing fun factors. It is considered as an approach to install game elements into non-game context activities~\cite{gamificationfornongame}. Deterding et al.~\cite{gamefulnessgamification} defined gamification as the use of design, elements, and characteristics of games in non-game context. They mapped gamification to serious games, playful design, and toys between two-axis, game/play, and whole/parts. Gamification was described relative to playfulness~\cite{playmatters}, which was considered similar to fun (e.g., ~\cite{4148841}). Huotari and Hamari~\cite{10.1145/2393132.2393137} presented a different perspective and defined gamification from the experiential nature of games as opposed to the systematic understanding of their elements.

\subsection{Fun in HCI}
To describe fun, particularly in the HCI field, the book {\it Funology} is recommended~\cite{blythe2004funology, blythe2018funology}. Several authors discussed enjoyment and HCI, and its role in usability. Hassenzahl and Tractinsky~\cite{uxagenda} discussed the user experience in HCI. They concluded their article by saying ``{\it From our perspective, one of HCI’s main objectives in the future is to contribute to our quality of life by designing for pleasure rather than for absence of pain. UX is all about this idea (pp. 95).}'' In fact, pleasure has been discussed to be an important human factor beyond usability~\cite{pleasureusability}. Draper~\cite{funsoftware} discussed cases where fun can be useful for software design. He argued that fun is obtained through the user's attention, and it is not what is always wanted. Fun can contribute in cases such as learning, but not in cases where the experience should become transparent. Kuts~\cite{playfului} performed a literature review for playful user interfaces, and provided an overview of interface features that can contribute to playfulness and enjoyment. Brandtz{\ae}g et al.~\cite{Brandtzaeg2018} discussed how enjoyment in human factors can be understood by referring to a previous model by Karsek and Theorell~\cite{karasek1990healthy}. They concluded that the design of enjoyable technology can result from user control and participation, with appropriate challenges, variation and multiple opportunities, and social opportunities in terms of co-activity and social cohesion. Blythe and Hassenzahl~\cite{semanticsoffun} discussed the semantics of fun. They considered whether synonymous words, such as pleasure, enjoyment, and fun refer to the same experience. They suggested that pleasure and fun differ in terms of their focus on an activity and a deep feeling of absorption. Furthermore, fun is considered in terms of distraction, whereas pleasure is of absorption. Therefore fun exists and can be a valuable point in HCI studies. Csikszentmihalyi~\cite{csikszentmihalyi1992flow} noted that art and leisure are not the only indicators for providing optimal experiences. These experiences can be observed in non-game HCI research. Furthermore, Thackara~\cite{thackara2001design} explained that technologies will not recede in the future because people enjoy them. Specifically, humans, particularly HCI researchers, are truly understanding the values of technologies because they are fun.

\begin{figure}[b]
  \includegraphics[width=1.0\columnwidth]{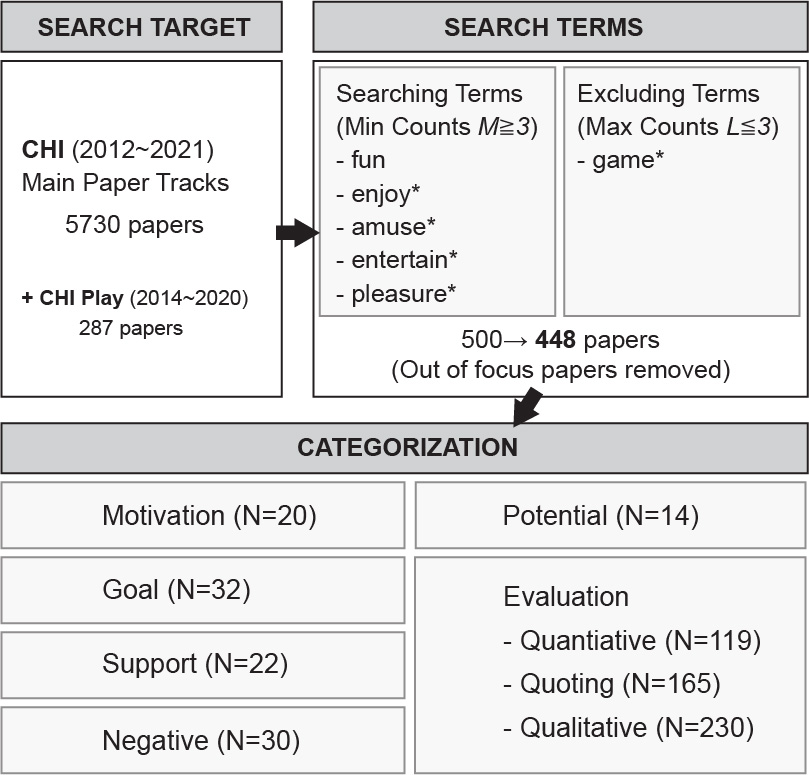}
  \caption{The flow of our searching procedure. We search on all papers presented in recent specific conferences. Several terms are used for the search, which narrows our dataset. We then categorize the results into six domains.}~\label{fig:method}
\end{figure}

\section{Method}
To explore the current understandings of enjoyment in HCI, we performed a literature review. Our method consists of a systematic searching phase, followed by screening and categorization (Figure~\ref{fig:method}).

\subsection{Database}
The source of our review consisted of 10 years of CHI proceedings (2012--2021) that focus on the main paper track. This resulted in a total of 5730 papers. We also went through the same process for the CHI Play proceedings (2014--2020) for reference (287 papers, note that CHI Play started at 2014). Similar to prior work with related methods~\cite{selfdethcigame}, we acknowledge that papers published at other relevant venues are excluded, however, we decided to focus on the flagship venue to work on a realistic number for the survey pool. 

\subsection{Searching Procedure}
\subsubsection{Search Terms}
To find research papers that use terms related to fun, and do not focus on game research, we prepared a python script to search for multiple terms. We noted that there are several synonyms to the term fun. Although previous studies showed that these synonyms could be used differently~\cite{semanticsoffun}, we extended our searching scope. This was inspired by Mekler et al.'s~\cite{enjoymentgames} work where they focused on the term {\it enjoy} and discussed the possibility of missing out terms such as {\it fun}. Additionally, because we assumed that non-game researches that use such terms are non-experts and the terms were used without careful distinguishing, we did not distinguish the synonyms in our procedure. For the synonyms used for searching terms, we referred to popular English dictionaries. In the Cambridge English Dictionary\footnote{https://dictionary.cambridge.org/dictionary/english/}, the term fun was descried as ``{\it pleasure, enjoyment, and entertainment.}'' Furthermore, in Lexico.com\footnote{https://www.lexico.com/} (an online dictionary powered by Oxford's English dictionary), the term was described as ``{\it enjoyment, amusement, or light-hearted pleasure}.'' Therefore, we resolved to use the terms fun, enjoy*, amuse*, entertain*, and pleasure* in the search procedure (* indicates wildcard to include close terms; e.g., enjoy, enjoyment, enjoyable, enjoyed etc.). To avoid searching through papers that focus on games, we also used the term game* for the searching procedure. An entire text search was completed for all target papers using these search terms. Two threshold values {\it M} and {\it L} were determined to narrow the searching results. They defined a minimum value of appearance of terms related to fun ({\it M}) and to limit the maximum value of appearance of game* ({\it L}). For our process, we found $M=L=3$ as a suitable threshold. Note that the search was suspended for sections beyond {\it REFERENCE} because it contains references that include terms in the name of the proceedings or the research title. Consequently, we obtained 500 papers. Interestingly, only one paper was retrieved from the CHI Play proceedings using this procedure reflecting the games focus of CHI Play.

We did not use the term play for our search because we noted that the term is often used in other contexts (e.g., play a role). Therefore, we mainly discuss work through the result retrieved regarding non-games context. We did not completely removed the term game* because it appears in context as reference in papers that do not have relation with game context.

\subsubsection{Screening Criteria}
For screening criteria, we particularly focused on (1) if fun was considered in the research, or (2) if the values or benefits of fun were discussed. In other words, papers in which the term fun was only used (1) to describe related works, (2) describe categories (e.g., entertainment as a genre), (3) describe contents (e.g., a fun app), (4) in a specific name, and (5) in a different context, were removed. This resulted in discarding 52 papers, and analyzing 448 papers. Through this process, we showed that our searching method contained target research papers and removed game-focused or highly related papers.

\subsubsection{Categorization}
The retrieved papers were initially categorized according to the features and tendencies of the usage of fun related terms. The papers were initially tagged with ambiguous words (a total of 29 words including words such as motivation and evaluation) to obtain necessary information to frame the categories. We then integrated and removed synonyms or related words for the final classification scheme. This resulted in six categories.
\begin{itemize}
  \setlength{\parskip}{0cm}
  \setlength{\itemsep}{0cm} 
\item Background (Motivation): explains why fun is required or important, i.e., it will contribute to motivation
\item Objective (Goal): aims to make something fun or treat the term as a purpose
\item Support (Proof): uses the term as a supportive element to strengthen the idea
\item Results (Evaluation): reports that an outcome of the study was fun, via quantitative or qualitative methods
\item Application (Potentials): proposes or mentions applicable use cases or describes potentials of a method
\item Negative: other cases where the term is used for negative cases
\end{itemize}

We also classified the papers according to their main characteristics (i.e., research type): (1) empirical studies, (2) design, (3) systems, and (4) hardware. In all cases, multiple categories were tagged in situations where a paper satisfied multiple schemes.

\section{Results}
Following the previously presented procedure, we categorized the papers into six domains. Although these categories included common understanding of enjoyment in game research, they are also common to be discussed in non-game context.

\subsection{Categories}
We obtained 20 papers indicating fun as a reason for motivation. Researchers treat fun as evidence where a user can use a system or to behave in a certain manner. One may argue their research approach can contribute to motivation by rewarding them that can be enjoyable~\cite{10.1145/3290605.3300235}. Enjoyment can lead to motivation for certain situations, such as children finding values for going out on trips~\cite{10.1145/3173574.3173810}, or for car applications owing to its entertainment~\cite{10.1145/2858036.2858336}. Enjoyment is often noted as an element to describe one's motivation for a certain concept.

In 32 papers, researchers set fun as a goal for their research or a certain activity. They note that designing an activity to be fun is an objective, and not a method. In studies that explore the user's demands, they may report that some systems can be used to simply entertain themselves~\cite{10.1145/2556288.2557273}. When researchers develop a system or tool, they set a hypothesis that their method can lead to enjoyable experiences~\cite{10.1145/3173574.3173930}. Researchers can consider a goal by hypothesizing and assessing enjoyment for their work.

In 22 papers, fun was used as evidence to show the value of their work or approach. In most cases, although it was premised that fun is valuable in some perspective, it did not provide explicit descriptions on contributions of fun classified works. For example, enjoyment can be described as a reason for why they chose a certain activity or method~\cite{10.1145/3411764.3445521}.

In 14 papers, fun was used as a potential of the work. Although the authors of those papers did not show that their claim was fun, they mentioned applications or visions that could lead to fun. These can result from qualitative feedback from participants in a study, suggesting further applications for improving entertaining concepts~\cite{10.1145/3025453.3025640}. In system papers, potential application scenarios can be introduced that are related to entertainment contents~\cite{10.1145/3290605.3300242}.

Occasionally ({\it N=30}), researchers use the term to describe null effects or clarify that it is beyond their scope; it is, or should not be fun. For example, it can be said that a mismatch could lead to unpleasant experiences in a certain context~\cite{10.1145/3290605.3300270}. In this case, the authors described the validity of design from the opposite perspective, i.e., one should design a system well to prevent negative effects.

Although not our main interest, we observed some trends in areas where terms regarding fun were used more as if it had presupposition of its value. In topics regarding accessibility or studies that work on people with any impairment or dementia ({\it N=24}), they often consider providing joy for target users. Hodge et al.~\cite{10.1145/3173574.3174088} showed a good example. They worked on virtual reality for people with dementia. The term enjoy* is repeatedly used to explain an approach that can effective by improving enjoyment. Similarly, researches on education have a popular trend ({\it N=4}). Clegg et al.~\cite{10.1145/3025453.3025987} considered learners with wearables and reported their observations and questionnaire results regarding enjoyment. It is a universal goal to accomplish positive emotions in the above cases. In other predictable cases, research on emotional studies ({\it N=8}), mainly use the term pleasure as a positive state. A paper from the CHI Play proceedings is in this class. It presented a study of novel auditory icons and recognition efficiency of emotions~\cite{10.1145/2793107.2793139}. We also find cases where arguments around humor are discussed, and it can contribute to improving the perception of task enjoyment~\cite{10.1145/3411764.3445068, humorrobot}. There seems to be an increase in work on virtual reality and related areas, for example, cases that note presence leads to an increase of enjoyment~\cite{10.1145/3411764.3445633}. Furthermore, discussions regarding joy and feeling of missing out are raised along with the contribution for presence~\cite{10.1145/3411764.3445183}.



\subsection{Measuring Fun}
From earlier discussions, we will give further details of the methods used to evaluate fun. We will provide examples so that one can refer to for their own research purposes. 


\subsubsection{Statements}
Questionnaires were used and reported to measure subjective enjoyment. Popular questionnaire statements were in forms such as: ``{\it I enjoyed \textasciitilde,}'' ``{\it \textasciitilde was fun,}'' ``{\it How enjoyable \textasciitilde,}'' and ``{\it \textasciitilde was enjoyable.}'' In unique cases, statements such as ``{\it Using the computer provides me with a lot of enjoyment}'' was used~\cite{10.1145/3173574.3173666}. In these cases, the term enjoyment was the most popular, followed by fun. The term like (``{\it I like to play with the 3D pen.}'') was used in one paper~\cite{10.1145/3290605.3300525} to measure enjoyment. 

\subsubsection{Standardized Questionnaires and References}
We observed several standardized questionnaires used or referred for the studies. Intrinsic motivation inventory (IMI)~\cite{imi} ({\it N=5}), $E^{2}I$ questionnaire~\cite{e2i} ({\it N=3}), creativity support index (CSI)~\cite{csi} ({\it N=3}), technology acceptance model 3~\cite{doi:10.1111/j.1540-5915.2008.00192.x} ({\it N=1}), cognitive absorption~\cite{agarwal2000time} ({\it N=1}), children's images of and attitudes towards curiosity (CIAC)~\cite{391005ccb0584c538c4d33b18d25c5d3} ({\it N=1}), and need for cognition~\cite{doi:10.1207/s15327752jpa4803} ({\it N=1}) were used. There was also a case where an author designed scale was used for auditory user interfaces~\cite{10.1145/3170427.3188659} and visual analog scales designed by referring to the CSI and the Usefulness, Satisfaction, and Ease of Use (USE) questionnaire. Arboleda et al.~\cite{10.1145/3411764.3445398} also used questions inspired from USE and Feeling of Flow~\cite{feelingflow}. To measure emotional responses (mainly pleasure), methods such as positive and negative affect schedule ~\cite{panas}, self-assessment manikin (SAM)~\cite{BRADLEY199449}, the affective slider (AS, an updated version of SAM)~\cite{10.1371/journal.pone.0148037}, affect grid~\cite{russell1989affect}, PAD (a scale to express all emotions from three dimensions: pleasure, arousal and dominance)~\cite{mehrabian1974approach, 10.1145/1520340.1520420, padmodel}, and Emocard~\cite{doi:10.2752/146069201789378496} were used. In other cases, Gupta et al.~\cite{10.1145/3313831.3376581} conducted measures by using opposing terms to evaluate user preference (enjoyable vs. efficient).

The flow theory~\cite{flowtheory} was often indicated in papers. Luria et al.~\cite{10.1145/3025453.3025786} studied interfaces for smart-home devices and referred to flow. The questionnaires were adapted from Zuckerman and Gal-Oz~\cite{ZUCKERMAN2013803} and removed 3 items owing to mismatches to their study. Bian et al.~\cite{10.1145/2858036.2858351} similarly referred to flow and emphasized the relevance of intrinsic interest and measured them by referring to Trevino and Webster~\cite{4dbce2202d644d888442bba612a02b4c}.

In other cases, Tausczik and Pennebaker~\cite{10.1145/2470654.2470720} referred to linguistic style matching~\cite{doi:10.1177/026192702237953} and engagement to observe enjoyment in the study. Kujala and Miron-Shatz~\cite{10.1145/2470654.2466135} adopted statements from psychological fields~\cite{MITCHELL1997421, doi:10.1111/1467-9280.03455}. Wang and Mark~\cite{10.1145/3173574.3173992} worked on a study for students and the influence of Facebook, and referred to two studies regarding engagement of learners for their measurement~\cite{Shernoff2014, shernoff2009cultivating}. Aitamurto et al.~\cite{10.1145/3173574.3174119} referred to Brooke's System Usability Scale (SUS)~\cite{brooke1996sus} and another earlier case~\cite{10.1145/2556288.2557270} that referred to the SUS. Similarly, Kim et al.~\cite{10.1145/3290605.3300316} referred to USE questionnaire that measures the usability~\cite{lund2001measuring}. In these scales, the authors typically refer to statements that measure the satisfaction, and consider them as enjoyability.

\subsubsection{How to Measure and Report}
From the results presented in the previous section, we provide a brief conclusion on the approach that can be acceptable. To measure subjective results of fun, it is popular to use subjective and direct questionnaires with a Likert scale with 5 or 7-point scale. Statistical analysis is not predominantly mandatory in this case, and results can be reported using stack bars. However, formally and precisely, using standardized scales such as IMI is acceptable, and statistical analysis should be performed (methodology will depend on the study design). Collective qualitative data are valid where interviews can be held directly with the participant. Quoting the participant's comment will be useful to provide actual thoughts and results from the participant's experience. Note that although we have noted popular tendencies from our review, such approaches are not predominantly definite. 

The topic regarding analysis methods can lead to complicated discussions, where there are often conflicting arguments in other research fields. There are documents that provide suggestions for using Likert-scale questionnaires, the methods of using and reporting ~\cite{likertmedian, souselikert}. For example, Jamieson~\cite{likertmedian} noted that ``{\it the intervals between values cannot be presumed equal}'' and therefore reporting them with means and standard deviation are inappropriate. In general, non-parametric methods should be used for statistical analysis (e.g., Wilcoxon tests)~\cite{souselikert}. However, many HCI researchers may report Likert-scale results by means and use parametric analysis methods. Some argue~\cite{10.1145/3025453.3025600} the validity of this approach by referring to articles that argue using parametric methods will not lead to wrong results~\cite{likertlaw}. These discussions recall the recent work by Vornhagen
et al.~\cite{statchiplay}, where they provided guides for statistical analysis and called for transparent research in CHI Play.

\section{Game and Non-Game Research}
We determined popular points and differences between non-game and game papers from our review. In this section, we present a comparison of our review with enjoyment in game-focused research as reported by Mekler et al. ~\cite{enjoymentgames}. Mekler et al. focused on quantitative studies and a method for evaluating enjoyment in digital games (87 papers were analyzed). Given that we used different criteria and focused on a different domain for the review, a comparison of the number of counts does not provide an appropriate meaning. However, we provide a comparison to find approximate tendencies.

We determined standardized methods used in game and non-game research. For example, methods such as IMI and SAM were used in both areas. Furthermore, although our review process considered more papers than Mekler et al., the frequency of usage of these types of methods was more limited. Conversely, it is interesting to note the use of methods, such as CSI, because it is a unique case used in the non-game context. In game research, other methods, including the Game Experience Questionnaire (GEQ), are used. However, it was not used in non-game research. Additionally, Mekler et al. reported the use of facial electromyography (EMG) to measure objective results that were not observed in non-game papers in our review. Interestingly, EMG methods were explored for user research and then employed for game research to gather quantitative data, thus this corresponded to a case wherein evaluation methods were recruited in game research from a non-game context and were used to learn to apply the methods for the evaluation of games ~\cite{nackebringing}. We consider this contradiction to occur because of the difference in focus on the measurement, where enjoyment can be key for game research and a supportive element for non-game research. Thus, game research is used more to evaluate the effect on enjoyment. Conversely, subjective questionnaires were similar, and statements such as ``{\it I enjoyed \textasciitilde}'' were often found in both areas. Although several studies referred to flow theory in game research, it is scarcely found in non-game research. This can be discussed to understand the experience of studies because the theory can be applied to games and non-game research. As an alternative to refer to optimal experience theories, non-game research discusses them in terms of usability. They refer to measures, such as the SUS, to discuss the usability and enjoyment of a system or design. There were certain tendencies for evaluation methods and discussion lenses in game and non-game research. It was popular to use subjective unformatted questionnaires to measure fun in studies. Given that Mekler et al. focused on quantitative studies, we have no evidence of qualitative strategies. However, we assume that the employment of interviews to collect direct comments is a typical strategy in both research areas.

To conclude, we believe that the term ``fun'' used in HCI research can be stated as follows: the state of being fun tends to be understood as a positive cognition that can lead to one's intrinsic motivation to use a certain interface or to stimulate interactivity rather than the experience itself. While fun in game research is analyzed from the perspective of player experience and is the most important goal of games~\cite{enjoymentgames}, HCI research tends to view game research from the lens of ``usability.'' In some studies,  fun is treated as an important core aspect of work, and enjoyment of a design/system is usually treated as sub-evidence. Prior literature~\cite{blythe2004funology, pleasureusability} proposed the use of HCI research from the perspective of usability and towards a wider set of problems related to fun. With respect to this point, it can be considered that HCI research is attempting to measure more around the enjoyment as well as usability. 

We observe an increase in the number of video game players~\cite{gamedata}, where the players consider the games as entertainment. Given that there are many benefits of playing video games~\cite{granic2014benefits}, it is necessary to examine research on games for enjoyable HCI. We believe that these concepts and applications are alternatives and related directions that HCI researchers can investigate. We should also be aware that positive playability (or usability) is not the only direction for improving enjoyment. Constraints in games can be designed and produce poor playability, which can also lead to enjoyment~\cite{10.1145/3377290.3377309}. This can be a direction for researchers where they can refer to game research to design enjoyable experiences with limited usability. Given that HCI and user experience research has been adopted and adapted in evaluation methods for game research, we can learn from various areas to understand the role of enjoyment in HCI.

\section{Grounding Fun in HCI}
We will now conclude the current understanding of fun in non-game HCI. Furthermore, we propose future directions to improve our understanding of the role of fun in HCI research based on our findings via categorization and comparison with game research.

\subsection{Today's Understanding}
In HCI research, researchers use terms related to enjoyment. However, given that enjoyment is subjective, it is challenging to find a universal base and understanding. To explain this, we categorized the cases that use enjoyment into six paradigms: motivation, goal, proof, evaluation, potential, and negative use. Specifically, motivation refers to the case wherein one explains that enjoyment will lead to motivation. This is universal to research areas, such as games or gamification, where motivation is a key element for design. Goal refers to the case wherein one explains enjoyment as an objective, i.e., the aim is to make something entertaining. Proof refers to the case where enjoyment is used as a supportive factor to strengthen an argument, and potential refers to the case when further applications are suggested based on fun-factors. Negative use cases refer to cases wherein one attempts to provide an argument from the opposite side, i.e., it is not valuable because it is not fun. Although these were predominantly based on an author's subjective understanding, enjoyment was used as a metric to evaluate systems, tools, hardware, and studies. However, most authors do not clarify their intention to measure enjoyment and consider them as a self-evident contribution. As we have noted that fun in HCI has been discussed from more than 30 years ago~\cite{fun1988, blythe2004funology, funsoftware, pleasureusability, uxagenda}, it turns out that these kinds of discussions and terms have disappeared over time. If we think about this tendency positively where fun is treated as a self-evident contribution, we may say that HCI researchers regard enjoyment as an evident value. To conclude, enjoyment is good and useful, and thus valuable to report the validity of an approach. However, this value is weakly discussed nowadays. The only exception is the case where researchers understand that enjoyment can lead to {\it motivation}, and in this case common knowledge is shared with gamification. This also may possibly be connected to the self-determination theory, and measurements of motivation, engagement, and experience~\cite{designmotivation}.

\subsection{Moving Forward}
It is important to understand the reason for asking questions such as ``{\it Is it fun?}''? Computing systems can contribute to humans from various aspects. They can be convenient tools or can even act as a machine that can lead to pleasure and entertainment. Furthermore, the measurement of fun in HCI can lead  to support arguments of a computing system that can be valuable from different aspects including usability and motivation. The contributions of enjoyment in HCI research can be understood from knowledge learned from games and play, including theories such as flow. We will now discuss directions for strengthening enjoyment validation in HCI research based on extant studies from related fields that are not limited to games and based on the analysis and discussions in this study.

\subsubsection{Extending Our Scopes}
Based on our literature review, we found discussions regarding motivation and usability that were linked with the role of fun. However, we can enhance our understanding from other perspectives that are often discussed in the field of HCI. Oulasvirta and Hornb{\ae}k~\cite{hciproblemsolve} discussed HCI research from the scope of problem-solving based on the philosophy of Larry Laudan~\cite{laudan1978progress}. Problem-solving is, in fact, a central concept of HCI and scientific research. Furthermore, fun can contribute in supporting problem-solving research in the HCI field. Matsumura et al.~\cite{shikakeology} proposed {\it Shikakeology}, which is a method for designing triggers for behavior change. They explained that combinations of psychological and physical triggers can affect human behaviors and lead to social or personal problem-solving. Similarly, Volkswagen proposed a concept termed as {\it The Fun Theory}, where it was argued that fun can change behavior to a better state~\cite{funtheorystairs}. An example of the Fun Theory is the piano staircase initiative. In this case, people are encouraged to use stairs as opposed escalators. The stairs were designed to play sounds like a piano as individuals walked on them. After the initiative was employed, 66~\% of the individuals used stairs. Current HCI researchers lack the point of view where fun can contribute to problem-solving. We did not observe papers in which the measured enjoyment was discussed from this perspective. Hence, based on the popular analog methods for promoting stair climbing~\cite{stairclimb} (e.g., visualizing calories on stairs), HCI researchers can investigate the application of their knowledge based on computer-based interfaces and apply them for behavioral changes.

Sustainable HCI (SHCI) is a key topic in the HCI field~\cite{nextsustainhci}, and recent articles suggest that sustainability is a challenge in the HCI field~\cite{sevenhci}. In fact, the recent CHI community investigated the concept of sustainability and discussed its progress and evaluation methods~\cite{changesustainhci, validsustainhci}. Although these researchers predominantly do not discuss sustainability based on factors such as enjoyment, we note that enjoyment can be considered as a contributing factor. With this respect, Fuchs and Obrist~\cite{hcisociety} noted that fun or enjoyment is a cultural design principle for a sustainable information society. Engagement is an essential topic in HCI~\cite{engagementhci, hciengageframe}. A framework of engagement can include a period of sustained engagement~\cite{engageframe}. Kevin Doherty and Gavin Doherty~\cite{engagementhci} presented a survey regarding engagement in HCI. Along with the results, they proposed that fun can be used as a strategy for designing engagement. Similarly, this was discussed from perspectives such as playfulness, humor, and gamification. Engagement can play a significant role as a trigger for changing human behavior, and can be used as a strategy for design~\cite{sustainabledfbc}. As discussed in the previous section, enjoyment can be considered as a key factor in behavioral change. There are significant correspondences between sustainability, engagement, and enjoyment. Therefore, enjoyment can contribute to SHCI, and HCI researchers should be able to discuss enjoyment as contributions to sustainability.

Several extant studies proposed research on systems, tools, and hardware by considering that fun from the perspective of consumption can be a practical point. In fact, the motivation for consumption can correspond to fun. It is known that consumption does not consistently lead to serious benefits. However, a product can correspond to something that leads to enjoyment, and can be used for distraction from anxiety~\cite{lai1995consumer}. Hence, consumption can be experience-based, such as going to bars and watching TV shows. Franke and Schreier~\cite{selfdesigneffort} described that people find value in self-designed products. The design process is important, where one plays effort and/or enjoys the procedure. Furthermore, one might feel positively (pleasure) when considering the purchase of items~\cite{neuralpurchase}. There is always an equilibrium between the pleasure of acquisition and pain of paying. This enables a discussion on understanding users as consumers that can help strengthen the values of HCI research. If a task is evaluated as fun, then it can be considered for stimulating an individual's possessiveness, which is valuable for mental health and pleasure. Thus, designing enjoyment contribute to opportunities for deployment. Additionally, by considering the fact that self-designed products are valuable, this type of perspective can be useful for conducting research related to personal fabrication or workshop-based studies.

The aforementioned perspectives can potentially benefit researchers in investigating areas for supporting their arguments on evaluating enjoyment. In additon to enjoyment or beyond enjoyment, the concept of celebratory technologies~\cite{celebhci} can be another direction. Grimes and Harper~\cite{celebhci} discussed directions for positive and exciting designs in HCI. In addition to considering technologies as functions, we should consider them as providing additional value in entertainment.

\subsubsection{Additional Directions}

Ultimately, HCI is fun. We found many papers indicating, measuring, or discussing aspects around fun, where we observed a strong interest in enjoyment among the community. In this section, we provide brief additional directions that we researchers should be aware of. These directions can lead to a better understanding of enjoyment measurements. First, we must be aware that humans/users can find interest in ``technology'' itself (e.g.,~\cite{thackara2001design}). We might simply think that a computing interface is fun because it is something new or original. Second, clarifying the validity of measuring fun can be especially important for ensuring that it measures the intended element of the study. An individual can find participation in a ``study'' itself as fun, and not the evaluated ``experience.'' Third, as we noted earlier, we must be aware that fun is not just about usability. As defined in popular games and play literature (e.g.,~\cite{halfreal}), an experience can be designed from the perspectives of rules, or more specifically, ``challenges.'' Poor usability leads to the potential of a {\it challenge}, which can lead to enjoyment, similar to that of games~\cite{10.1145/3377290.3377309}. However, these elements were not clearly evident in our review.

If we revisit the early literature of fun and usability~\cite{fun1988, blythe2004funology, funsoftware, pleasureusability}, we may observe that some of the directions were already indicated at that time. For example, Caroll and Thomas~\cite{fun1988} noted that if a system is more fun, it can be a reason for it to be used more and to promote learning. These are related to our findings around education (Section 4.1) and our discussions around sustainability. Through our review, we only found a limited number of papers working on evaluating enjoyment. However, referring to Draper~\cite{funsoftware} who said that fun is obtained through the user's attention and is not always required, it may be correct not always to discuss enjoyment in our research. In addition, it may include cases where the enjoyment is becoming transparent and unnotable to the users. This can be one direction and goal for us, however, we still believe that discussing the reasons and motivations around enjoyment in HCI will lead to better understandings and contribute to passing down the background to the future. Furthermore, as Hassenzahl and Tractinsky~\cite{uxagenda} argued that one of the HCI's main objectives is to contribute to the quality of life through designing pleasure, designing a pleasurable experience can be an important direction. Celebratory technologies~\cite{celebhci} can be a similar idea where they discuss positive design through food, as well as the delight of food and eating~\cite{10.1145/3411764.3445218}. 

Despite the differences in understanding by current researchers, it is not essential to criticize or reject. We do think that we can work on better research papers through improved writing. Simultaneously, it can be the role of readers where they can fill the margin of missing validation of enjoyment through their knowledge. We expect that this paper to help the understandings of researchers from both perspectives of writers and readers, and to provide higher confidence in evaluations and discussions regarding enjoyment.

\section{Limitations}
We now discuss some limitations of our study and other potential factors that can contribute to extending our study.

\subsection{Terms and Tracks}
As mentioned earlier in this paper, the terms related to fun can be used in slightly different contexts~\cite{semanticsoffun}. In our review, we did not notice a significant difference in the usage of the terms {\it fun} and {\it enjoyment}. However, there were tendencies for {\it pleasure} and {\it entertain}. The term {\it pleasure} was often used to measure the psychological effects, and the term {\it entertain} was often used to indicate specific contents. Our study can be extended via a deeper examination of the differences in usage of such terms. Furthermore, based on our review, we recognized that there are several terms that are closely related to the terms used in our search procedure. The following terms were often noticed in the reviewed papers: {\it interest, joy, engagement, happiness, funny, pleasant, arousal, satisfaction, excitement}, and {\it playful}. We assume that by using these words in the criteria of review, we can extend our study for a deeper understanding. Ryokai et al.~\cite{10.1145/3173574.3173932} captured laughter to seek happiness, and this also discussed by using the term enjoyment. Given that we referred to engagement in this paper, focusing on this point will be particularly be important for further research. There are also cases where enjoyment is described with the term relaxation~\cite{poels2012}.

The alt.chi track at CHI conferences can be an interesting area of exploration. Owing to the nature of where the alt.chi track invites critical papers and encourages abnormal presentation formats, we observe entertaining ideas. For example, ideas such as stupidity, were introduced for creative thinking~\cite{10.1145/2559206.2578860}. Furthermore, in Lickable Cities~\cite{10.1145/3170427.3188399}, taste was explored via licking of urban spaces. The authors stated that ``{\it we hope to entertain, disgust, and/or inspire readers with our absurdist research endeavor.}'' Thus, the authors from this track aim to present entertaining contents through their paper. Therefore, fun can also be a desirable element for attracting for criticisms and social discussions, and we can consider potential discussions regarding research methods. We introduced several potential areas where researchers can think of extending their scope when dealing with enjoyment. The discussed areas are introduced based on the author's knowledge that was not found through the literature review. Therefore, we expect other research areas related to our topic, such as speculative design, design fictions and humor. 

\subsection{Systematic Search}
Given that our review method was systematic, our criteria can significantly affect our results. Although fun was discussed in the scope of games and play in the earlier sections, the term play was not used in our search criteria. Therefore, our results can be slightly biased with respect to the context of play.

In this study, another problem arises from our approach owing to the session names of the conference that are typically included in the header of the published paper. For example, in CHI 2017, there was a session called {\it novel game interfaces}. In such a case, our automatic searching procedure counts the word game in the header of the paper, even if the term game was not included in the main content. In the paper by Lopes et al.~\cite{10.1145/3025453.3025600} (presented in CHI 2017, Novel Game Interfaces session), game was found 12 times (excluding pages beyond {\it REFERENCES}); 10 times in the header and 2 times in the main content. Given that we set the maximum count of game as 3 in our search criteria, without the counts in the header, this paper should have been included in our search. This becomes particularly difficult problem because all papers do not successfully embed a table of contents information into the published PDF document. Unfortunately, other related issues can occur due to our automatic searches (e.g., bias towards page lengths and, different numbers depending on the word count criteria). However, it was necessary to control our review to ensure that it was traceable and retain the amount of human effort to be realistic. 

\section{Conclusion}
Enjoyment is an important factor that yields several benefits, and thus it is used as a metric in several studies on HCI research. To explain the current understanding and treatment of enjoyment by the CHI community in a non-game context, we performed a review and categorized the findings into several domains. Furthermore, we evaluated methods that are used to measure fun. Some researchers argue that enjoyment can lead to motivation, but we still think that HCI researchers should be aware of other potential aspects of the role of enjoyment. Beyond the discussions for enjoyment and the relation with usability, we argue that we are still missing perspectives and suggest fun should be used as a metric because it can be valuable in many contexts. Thus, there are more opportunities where we can further strengthen the concept of enjoyment in HCI. A computing system that is {\it fun} can lead to positive effects. We can find pleasure from various interactions that occur in HCI, and this contributes to an enhanced experience.


\bibliographystyle{ACM-Reference-Format}
\balance
\bibliography{sample-base}

\end{document}